\documentclass[10pt,a4paper,twocolumn]{article}

\usepackage{lmodern}          
\usepackage[T1]{fontenc}
\usepackage[utf8]{inputenc}
\usepackage[margin=2.5cm]{geometry}
\usepackage{graphicx}
\graphicspath{{figures/}}
\usepackage{booktabs}
\usepackage{siunitx}
\usepackage{amsmath}
\usepackage{microtype}
\usepackage[numbers,sort&compress]{natbib}
\usepackage[colorlinks,allcolors=blue!60!black]{hyperref}
\usepackage{xcolor}
\usepackage{stmaryrd}
\usepackage{authblk}

\DeclareSIUnit{\rydberg}{Ry}
\providecommand{\gibi}{Gi}\providecommand{\byte}{B}

\newcommand{\Nseed}{16}
\newcommand{\Ncorpus}{298}
\newcommand{\Noa}{135}
\newcommand{\Nfulltext}{69}
\newcommand{\Naudited}{57}
\newcommand{\Nexcludednotpenta}{5}
\newcommand{\Nextractedok}{62}
\newcommand{\Narxiv}{41}

\newcommand{\Nforbidden}{36}

\newcommand{\Nreproducible}{19}
\newcommand{\PctReproducible}{33.3}
\newcommand{\PctMeanReported}{68.3}
\newcommand{\NpreprintPapers}{17}
\newcommand{\NpreprintComplete}{10}
\newcommand{\NpublisherPapers}{40}
\newcommand{\NpublisherComplete}{9}
\newcommand{\PctPreprintComplete}{59}
\newcommand{\PctPublisherComplete}{22}
\newcommand{\PreprintGapDiffPp}{36}
\newcommand{\PreprintGapFisherP}{0.013}

\newcommand{\Nfields}{475}
\newcommand{\NfieldsGrounded}{473}

\newcommand{\PctSupported}{95.3}
\newcommand{\Nvaluechecked}{169}

\newcommand{\Nclaimscleared}{51}
\newcommand{\Pctclaimscleared}{16.4}

\newcommand{\Model}{qwen3:4b}
\newcommand{\GPU}{NVIDIA RTX~3050 Laptop (4\,GiB)}
\newcommand{\NlayersResidentMin}{23}
\newcommand{\NlayersResidentMax}{37}
\newcommand{\NlayersTotal}{37}

\newcommand{\NlabelPapers}{25}
\newcommand{\NlabelJudgements}{201}
\newcommand{\NlabelStubs}{2}
\newcommand{\PctRecall}{67.3}
\newcommand{\PctRecallLo}{59.8}
\newcommand{\PctRecallHi}{74.0}
\newcommand{\PctPrecision}{95.7}
\newcommand{\PctPrecisionLo}{90.3}
\newcommand{\PctPrecisionHi}{98.1}

\newcommand{\PctRecallEight}{69.8}
\newcommand{\NcodeCorrect}{8}
\newcommand{\NcodeChecked}{10}
\newcommand{\Nconversionlost}{6}
\newcommand{\PctConversionLoss}{6.7}

\newcommand{\EvalRtxPresP}{95.7}
\newcommand{\EvalRtxPresR}{67.3}
\newcommand{\EvalRtxPresF}{79.0}
\newcommand{\EvalRtxAbsP}{36.5}
\newcommand{\EvalRtxAbsR}{86.1}
\newcommand{\EvalRtxAbsF}{51.2}
\newcommand{\EvalRtxMcc}{0.414}
\newcommand{\EvalTfourPresP}{96.3}
\newcommand{\EvalTfourPresR}{78.8}
\newcommand{\EvalTfourPresF}{86.7}
\newcommand{\EvalTfourAbsP}{47.0}
\newcommand{\EvalTfourAbsR}{86.1}
\newcommand{\EvalTfourAbsF}{60.8}
\newcommand{\EvalTfourMcc}{0.530}
\newcommand{\EvalFfsixteenPresP}{99.2}
\newcommand{\EvalFfsixteenPresR}{73.9}
\newcommand{\EvalFfsixteenPresF}{84.7}
\newcommand{\EvalFfsixteenAbsP}{44.9}
\newcommand{\EvalFfsixteenAbsR}{97.2}
\newcommand{\EvalFfsixteenAbsF}{61.4}
\newcommand{\EvalFfsixteenMcc}{0.560}

\newcommand{\EvalPctMinority}{18}

\newcommand{\PctGroundedPre}{93.3}     
\newcommand{\Nselfneg}{30}             
\newcommand{\PctSelfneg}{5.7}
\newcommand{\PctLocaliserNarrow}{58.2} 
\newcommand{\PctLocaliserWide}{71.1}
\newcommand{\NdevExtractions}{509}
\newcommand{\NverifRecords}{12}
\newcommand{\NtranslatorCalls}{17}
\newcommand{\NconvCalls}{24}
\newcommand{\NcriticCalls}{13}
\newcommand{\NcriticCollapse}{3}
\newcommand{\NcriticToolCalls}{48}
\newcommand{\NtranslatorSets}{11}
\newcommand{\NallAgentCalls}{54}
\newcommand{\NcriticChanged}{4}
\newcommand{\NdevCampaigns}{ten}
\newcommand{\NdevTokens}{4.7}        

\newcommand{\NcacheShared}{8}
\newcommand{\CacheCellMax}{0.0052}

\newcommand{\NfieldsPassOne}{435}
\newcommand{\NfieldsPassTwo}{440}
\newcommand{\NfieldsPassThree}{439}
\newcommand{\NzeroPassOne}{18}
\newcommand{\NzeroPassTwo}{20}
\newcommand{\NzeroPassThree}{20}
\newcommand{\NzeroUnion}{17}
\newcommand{\Nrecovered}{35}
\newcommand{\Ncollapsedonce}{21}
\newcommand{\Ncollapsedalways}{17}
\newcommand{\Nrescued}{4}

\newcommand{\NclaimsTierTwo}{8}
\newcommand{\NprototypesTierTwo}{3}
\newcommand{\NpapersTierTwo}{4}
\newcommand{\MedianDeviation}{1.1}
\newcommand{\MAEDeviation}{0.110}
\newcommand{\BiasDeviation}{0.090}
\newcommand{\MAREDeviation}{2.3}
\newcommand{\NwithinTwoPct}{6}

\title{\bfseries Evaluating a 4B open-weights local LLM for agentic DFT workflows: a literature reproducibility audit}

\author[1]{Shambhu Bhandari Sharma\thanks{shambhu.sharma.22@ucl.ac.uk}}
\affil[1]{University College London, Department of Earth Sciences,
London WC1E 6BT, United Kingdom}

\date{\today}

\begin{document}
\twocolumn[
  \begin{@twocolumnfalse}
    \maketitle

\begin{abstract}
\noindent
Agentic workflows in materials science relying on hosted commercial models face severe reproducibility, economic, and data-privacy constraints. To explore fully local agentic science, this work evaluates an open-weights \texttt{\Model{}} model executing an autonomous scientific pipeline across varying hardware constraints. Applied to pentagonal two-dimensional materials, the system extracts parameters from unstructured text, translates them into density functional theory (DFT) inputs, and drives simulations to convergence under a strict neurosymbolic architecture where agents propose and deterministic code disposes. The workflow is guarded by verbatim text grounding and multi-pass inference unions to counteract hardware-induced structural collapse. Evaluated against \NlabelJudgements{} expert judgements, the extractor achieves \PctPrecision\,\% precision (95\% CI \PctPrecisionLo--\PctPrecisionHi\,\%) and \PctRecall\,\% recall (\PctRecallLo--\PctRecallHi\,\%), ensuring extracted parameters are strictly factual. However, precision identifying absent parameters does not exceed \EvalTfourAbsP\,\%, establishing that the measured omission rate constitutes a loose upper bound on true literature incompleteness. Across three hardware configurations, complete GPU residency governs extraction quality more fundamentally than weight or cache precision, raising Matthews' correlation from \EvalRtxMcc{} to \EvalTfourMcc{} at fixed quantisation, and to \EvalFfsixteenMcc{} with an unquantised cache. A corpus-scale audit indicates only \Nreproducible{} (\PctReproducible\,\%) of the \Naudited{} studies are reproducible in principle, reporting every method parameter needed to re-initialise the calculation. Driven to convergence, the workflow reproduces published lattice constants with a mean absolute relative error of \MAREDeviation\,\% where the relaxed structure retains its prototype, establishing that lightweight open-weights models can reliably drive autonomous agentic workflows when bounded by deterministic code gates.
\end{abstract}

\vspace*{1.5em}
  \end{@twocolumnfalse}
]

\section{Introduction}
\label{sec:intro}

Large language model (LLM) agents are already driving autonomous workflows across computational chemistry and materials science. Recent pipelines span the entire research lifecycle, from coupling expert tools for chemical planning and executing robotic synthesis~\cite{m2024augmenting, boiko2023autonomous, szymanski2023autonomous}, to orchestrating density functional theory (DFT) calculations through hierarchical multi-agent frameworks~\cite{wang2025dreams, yang2026autodft}. Within materials informatics, transformer-based architectures have firmly established the viability of extracting composition-property records at corpus scale~\cite{swain2016chemdataextractor, roy2026comproscanner}. These systems depend on grammar-constrained decoding to ensure parseable output and strict grounding techniques to verify generated statements against source texts~\cite{tang2024minicheck}, and must confront known reproducibility issues with non-deterministic floating-point math during inference~\cite{yuan2025understanding}.

However, almost all of these agentic systems share a critical dependency: they are driven by hosted, closed-weight frontier models. This reliance on commercial APIs introduces significant operational and epistemological vulnerabilities. A workflow tied to a commercial key ceases to run when the endpoint is repriced, the model is reversioned, or the key expires, meaning long-term reproducibility is never guaranteed. Furthermore, metered inference is billed per token, making iterative debugging prohibitively expensive. Frontier inference also carries a measurable energy cost per query~\cite{jegham2025hungry}. Crucially, much computational chemistry operates on isolated networks where proprietary data cannot be sent to third-party endpoints, rendering cloud-hosted models unusable.

Fortunately, capability on agentic tasks appears to depend more on interface design than raw parameter count. Recent studies demonstrate that tool-augmented smaller models can outperform larger unconstrained baselines~\cite{zywot2026can}. A 4B retrosynthesis planner surpasses a 7B reasoning baseline given a constrained tool interface~\cite{zhu2026retroagent}, and 7-8B chemistry agents on a consumer GPU have matched datacentre-hosted equivalents~\cite{mcnaughton2024cactus}. 

While the $\Delta$-project established that independent DFT codes agree closely given identical inputs~\cite{lejaeghere2016reproducibility}, it presupposed that methodological parameters are completely and transparently reported in the source text. Whether computational literature, spanning both peer-reviewed articles and preprints, provides sufficient methodological transparency to enable autonomous simulation recomputation across an entire corpus remains completely untested in density functional theory.

This work applies a fully open-source, local agentic workflow to address this critical gap: how often does a first-principles study report sufficient parameters to permit independent replication? Every component runs locally and unmetered, on an open-weights \texttt{\Model{}} model~\cite{yang2025qwen3} on a single local GPU via Ollama~\cite{ollama2023} and the SIESTA electronic structure package~\cite{soler2002siesta}. Bounding the model's autonomy is the central architectural constraint: agents select among named options rather than free-form arguments, and every proposal passes a deterministic code gate before reaching a calculation or tally~\cite{doshi2026towards}. 

The contributions are threefold: a neurosymbolic architecture spanning extraction, input translation, and autonomous execution; a taxonomy of silent failure modes including head-truncated prompts and structural collapse under incomplete GPU residency~\cite{barnett2024seven}; and a case study auditing \Naudited{} first-principles studies of pentagonal lattice-based~\cite{zhang2015penta} two-dimensional materials, recomputing published lattice constants against fresh SIESTA calculations.

\section{Methods}
\label{sec:methods}
\begin{figure*}[t]
\centering
\includegraphics[width=\textwidth]{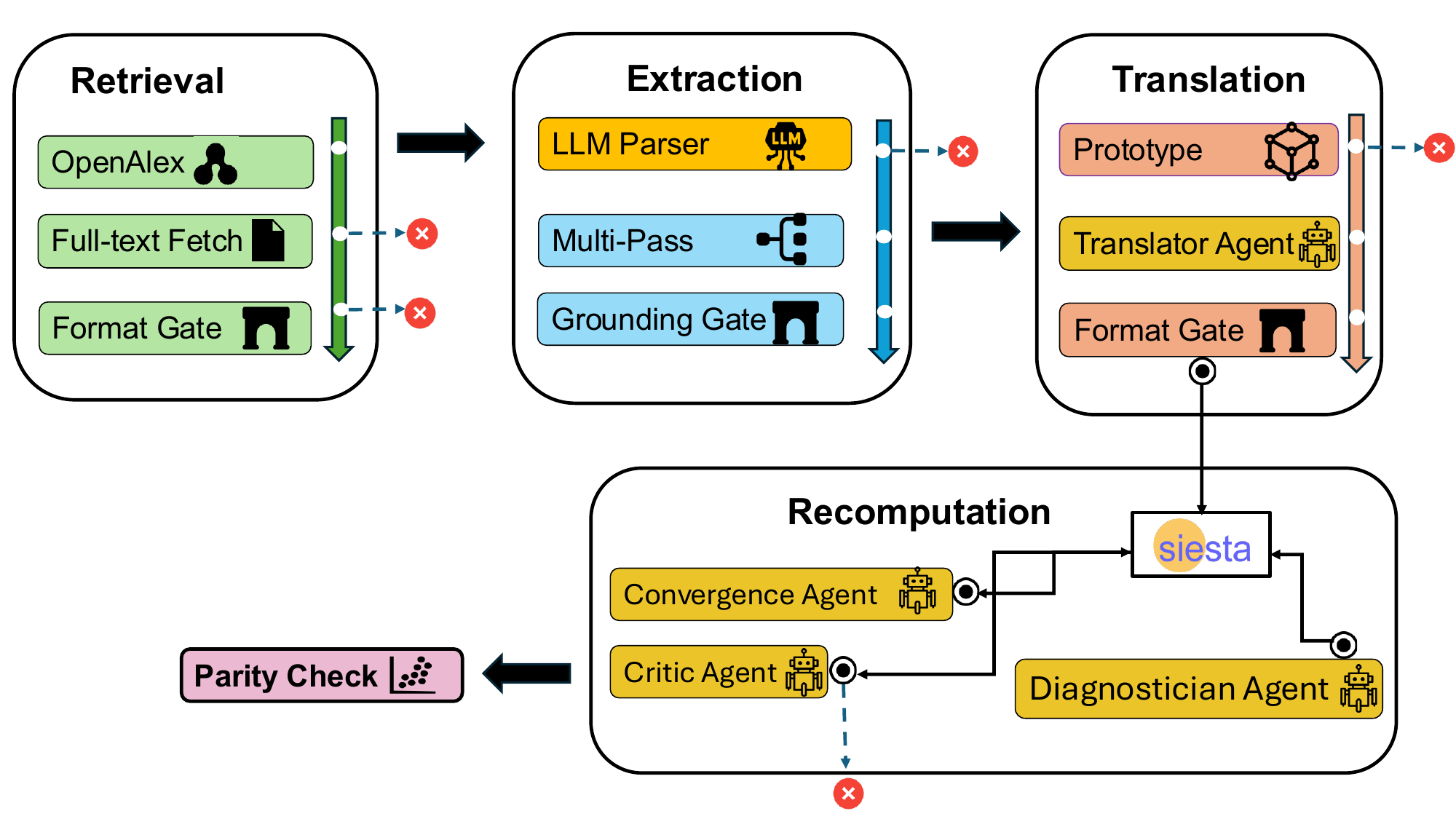}
\caption{The autonomous audit pipeline. Retrieval fetches and screens OpenAlex full texts. Extraction parses and deterministically grounds parameters across three passes. Translation maps extracted claims and topological prototypes to SIESTA inputs. Recomputation drives physical simulations to convergence, monitored by diagnostician and critic agents. The workflow concludes with a parity check against the literature. Yellow boxes denote LLM agents; every other box is a deterministic code step. Red crosses indicate discarded records.}
\label{fig:workflow}
\end{figure*}

This section details the system architecture and hardware constraints governing an automated workflow that spans four sequential phases, from document acquisition to physical simulation (Figure~\ref{fig:workflow}).

\subsection{System architecture and constraints}
\label{sec:agents}

\paragraph*{Hardware constraints and model quantisation.} All extraction and translation agents run \texttt{\Model{}} locally via the Ollama inference server~\cite{ollama2023} (backed by \texttt{llama.cpp}~\cite{gerganov2023llamacpp}) on a \GPU{}, the constrained deployment target the method is designed for. To execute the model within the strict VRAM capacity of this hardware class, both weights and the attention key/value (KV) cache are quantised. Weights are statically quantised to \texttt{Q4\_K\_M}, and, on the \SI{4}{\gibi\byte} GPU, the KV cache is additionally quantised to 4-bit (\texttt{q4\_0}) to accommodate the sequence length. Residency is not guaranteed on a GPU of this capacity: the placement of individual layers between device and host is determined by the inference server at load time and is not fixed by configuration. To decouple the two quantisation axes from this uncontrolled residency condition, the  pipeline is additionally evaluated on a Tesla~T4 (\SI{15}{\gibi\byte}), which accommodates the model in full and permits weight and cache precision to be varied independently (Section~\ref{sec:results-stability}).

\paragraph*{Inference parameters and grammar constraints.} Operating within these physical constraints, all agent generations use greedy decoding ($T = 0$) with a generation limit of 2048 tokens, a cap on output; internal reasoning (`thinking') is explicitly disabled to prevent models from expending token budgets on chain-of-thought sequences that the grammar discards. The context window dynamically accommodates the document length; truncated prompts are retried at double capacity, and those exceeding the 32{,}768-token ceiling are recorded as failures. Output is strictly constrained at the logits level using a grammar compiled directly from the target JSON schema (GBNF). Two schema properties are lost during compilation and require explicit programmatic compensation. First, semantic field descriptions are stripped, necessitating their re-injection into the zero-shot prompts. Second, fields carrying default values are not inherently marked as required. All schema fields are explicitly declared as \texttt{required} prior to compilation; for the extraction schema, where this forced the model to emit a null object for every parameter, only the fields determining reproducibility are required. String length and pattern constraints, which the grammar compiler rejects, are removed from the compiled schema and enforced on the returned object instead.

\paragraph*{Neurosymbolic architecture: agents propose, code disposes.} Because syntactic schema compliance does not guarantee factual correctness, the pipeline enforces a strict neurosymbolic hierarchy: agents propose, and deterministic code disposes. To prevent arbitrary code execution, each agent is restricted to a named allow-list of read-only tools, and tool arguments are programmatically bound. The agent selects a routine by name, arguments are populated from the current verification state, and tool execution rounds are strictly capped. Every returned object passes a deterministic gate: Pydantic~\cite{colvin2017pydantic} validation of the schema and length constraints withheld from the grammar, followed by explicit range and vocabulary checks in Python. Agent proposals failing these checks are corrected or refused, and the intervention is logged; values extracted from a paper are instead flagged and recorded unaltered, since an out-of-range value may be an extraction error or the paper's reported choice, and the two are separate findings. Each agent call is written to a JSONL audit log recording what the agent proposed, what was finally applied, and which gate fired on any difference, from which override counts per agent are tallied.

\subsection{Phase 1: Literature retrieval}
\label{sec:corpus}

\paragraph*{Metadata retrieval.} The corpus is constructed via a three-stage pipeline, beginning with the OpenAlex API~\cite{priem2022openalex}, selected over proprietary databases such as Scopus for its keyless, freely redistributable access and its indexing of preprints alongside publisher records. A curated set of \Nseed{} seed terms is applied to titles and abstracts: nine structural terms targeting topological families (e.g., ``Cairo tiling'') and seven heavily studied compositions (e.g., \texttt{penta-SiC2}), retrieving records whose abstracts designate solely the chemical formula.

\paragraph*{Filtration and deduplication.} Each record passes four deterministic filters. It must be an article or preprint, and its DOI must not match patterns indicating supplementary information. Furthermore, it must contain a first-principles indicator in its title or abstract (e.g., ``first-principles'', ``density functional'', or a named DFT code), and match a pentagonal structural pattern. Surviving records are deduplicated (grouped by arXiv identifier, otherwise by normalised title), preferring non-arXiv DOIs and higher citation counts. The published DOI is retained as the primary key, with the OpenAlex identifier as fallback; arXiv identifiers and open-access URIs are preserved to facilitate access to paywalled full text.

\paragraph*{Full-text acquisition and lexical validation.} Full-text acquisition prioritises institutional repositories and arXiv. Documents are parsed natively (PDFs via PyMuPDF~\cite{artifex2024pymupdf}, HTML via BeautifulSoup~\cite{richardson2004beautifulsoup}) into raw text arrays. To ensure only verified, full-length manuscripts proceed, a document is accepted if at least four of nine structural markers (e.g., ``abstract'', ``conclusion'') are present; a document shorter than 12{,}000 characters must additionally contain at least three methodological markers (e.g., ``cutoff'', ``pseudopotential''), and documents yielding fewer than 2{,}000 characters are discarded. Supplementary information is never fetched; a parameter documented solely in the supplementary material is recorded as absent (Section~\ref{sec:discussion}).

\paragraph*{Deterministic subsampling.} Where a subset is necessary for evaluation, it is drawn deterministically via a SHA-256 hash of each paper's identifier. This mechanism yields a strictly reproducible sample uncorrelated with year, venue, or citation count.

\subsection{Phase 2: Parameter extraction}
\label{sec:eval}

For each retrieved manuscript, the extraction agent evaluates a schema of twenty distinct computational method parameters, returning for each a boolean indicating presence, the extracted numerical value or string, and the verbatim sentence supporting the extraction. A paper is defined as \emph{reproducible in principle} if it reports every parameter required to re-initialise the calculation. Because the required parameter set is code-dependent (e.g., a plane-wave cutoff and a real-space mesh cutoff are not interchangeable), these specific requirements are defined deterministically in the pipeline's configuration.
Formally, for a paper $p$ reporting code $c(p)$, the scored set is
\begin{equation}
S(p) = R_{\mathrm{univ}} \cup R_{\mathrm{2D}} \cup B\bigl(c(p)\bigr) \cup \{\texttt{code}\},
\label{eq:required}
\end{equation}
where $R_{\mathrm{univ}}$ denotes universal DFT settings (\texttt{functional}, \texttt{pseudopotential}, \texttt{k\_mesh}, \texttt{force\_threshold}, \texttt{energy\_threshold}), $R_{\mathrm{2D}}$ denotes two-dimensional monolayer constraints (\texttt{vacuum\_spacing}), $B(c) = R_{\mathrm{PW}}$ (\texttt{plane\_wave\_cutoff}) for plane-wave codes, $R_{\mathrm{NAO}}$ (\texttt{mesh\_cutoff}, \texttt{basis\_size}) for numerical-atomic-orbital codes, and $\emptyset$ where no code is stated. The reported fraction is
\begin{equation}
f(p) = \frac{\bigl|\{\, s \in S(p) : \mathrm{rep}_s(p) \,\}\bigr|}{|S(p)|},
\label{eq:fraction}
\end{equation}
where $\mathrm{rep}_s(p) \in \{0, 1\}$ indicates whether parameter $s$ is reported in paper $p$, and $p$ is reproducible in principle exactly when $f(p) = 1$.

\paragraph*{Full-context information extraction.} To preserve global context for parameter disambiguation, the full text (excluding the bibliography and truncated to a 120{,}000-character limit) is supplied to the language model in a single prompt context, strictly circumventing retrieval-augmented generation.

\paragraph*{Multi-pass extraction and grounded union.} Given the variance of greedy decoding, full-context inference is executed in three independent passes. Outputs are combined via a grounded union rather than a majority vote to mitigate whole-paper collapse. Writing $S_k$ for fields asserted in pass $k$, $e_x$ for the supporting textual evidence quoted for field $x$, $d$ for the source document text, and $G$ for the evidence gate, the validated set of a pass ($V_k$) and the shipped record ($E$) are:
\begin{equation}
V_k = \{\, x \in S_k : G(e_x, d) = 1 \,\},
\qquad
E = \bigcup_{k=1}^{K} V_k,
\label{eq:union}
\end{equation}
with $K = 3$. This monotonic union ensures a collapsed pass ($V_k = \emptyset$) withholds evidence rather than falsely voting for absence, and ensures widening coverage cannot admit ungrounded claims. What the union cannot repair is misattribution, since a correctly quoted sentence assigned to the wrong field satisfies $G$ in every pass.

\paragraph*{The deterministic grounding gate.} Extraction is governed by a strict, one-directional evidence gate: asserted parameters must supply verbatim textual evidence; absent claims are never challenged. Prior to verification, quotes are normalised ($\mathcal{N}$) by stripping model framing, lowercasing, and converting hyphens to spaces. Writing $w$ for the words of the normalised quote and $n = 6$, evidence $e$ is accepted against document $d$ when:
\begin{equation}
G(e,d) =
\begin{cases}
1, & \mathcal{N}(e) \subseteq \mathcal{N}(d),\\[2pt]
1, & |w| \ge n \;\wedge\; \exists\, i : w_{i:i+n-1} \subseteq \mathcal{N}(d),\\[2pt]
0, & \text{otherwise.}
\end{cases}
\label{eq:ground}
\end{equation}
If $G(e,d)=0$, the boolean is reset ($\mathrm{rep}_s \leftarrow 0$) and the override logged.

\paragraph*{Unit-aware value support.} Where the grounding gate verifies the existence of the quoted sentence, a secondary value-support diagnostic evaluates whether any reported numerical value mathematically follows from that text within a relative tolerance. Operating strictly as an observational diagnostic, this unit-aware check does not override the primary boolean extraction; the value-grounding gate described below does. To correctly evaluate standardized numerical values, the algorithm applies deterministic, unit-aware conversions (e.g., \si{\rydberg}\,$\to$\,\si{\electronvolt}) via a predefined unit conversion mapping, decoupling semantic accuracy from syntactic formatting.

\paragraph*{Numeric value grounding.} A second gate governs the
numerical content of a claim. Where the evidence gate establishes that a quoted sentence
occurs in the source, the value gate requires that the asserted magnitude itself occur
verbatim within the document. A value that cannot be so located is cleared while the
surrounding claim is retained, so that an ungrounded magnitude is removed without
discarding the evidence that the parameter was discussed. The gate is deterministic, and consequently reproducible without re-invoking the
model; it is applied at extraction time and, for corpora predating its introduction,
retrospectively under identical criteria.

\paragraph*{Quantifying systematic error: precision versus recall.} Finally, to evaluate the extractor's baseline recall rate against these automated gates, an expert annotator produced hand-labelled ground truth for a targeted subset of core schema fields across a subset drawn from the hash-sampled evaluation set (retaining only viable manuscripts and applicable judgements). The annotator labelled from the published PDFs rather than from the converted text, so that a parameter lost in conversion could be separated from one the authors never reported; the labels are a single expert's judgements, and no second-rater agreement statistic is
claimed. A second batch of papers was labelled from the retrieved text itself, recording
only the reported/not-reported decision, which is the sole input to precision and recall.

\subsection{Phase 3: Input translation}

\paragraph*{Initialisation and topological prototypes.} Because author-provided coordinates are universally absent, structures are generated exclusively from recorded topological prototypes and scaled via empirical covalent radii. Compositions lacking a known pentagonal prototype are skipped. These unrelaxed geometries are passed directly to SIESTA~5.4.2, enforcing a \SI{20}{\angstrom} out-of-plane vacuum. The calculations employ scalar-relativistic norm-conserving ONCVPSP pseudopotentials in PSML format alongside a baseline double-$\zeta$ polarised (DZP) numerical atomic orbital basis.

\paragraph*{The Translator agent and override guardrails.} The DFT calculation is configured by the Translator agent, which maps extracted parameters onto SIESTA input directives. Consistent with the neurosymbolic architecture, the agent's proposals are strictly governed by thirteen deterministic gates. Rather than halting execution or prompting the agent, these gates programmatically override malformed proposals with physically consistent fallback parameters and log the intervention. For example, the gates force the out-of-plane $k$-grid component to unity for a 2D monolayer, enforce a minimum $9 \times 9 \times 1$ mesh for sub-unity grids, balance in-plane grid components differing by more than a factor of two, and reset mesh cutoffs outside the physical bounds of \SIrange{50}{2000}{\rydberg} to a default of \SI{350}{\rydberg}. Furthermore, the gates enforce semantic boundaries on simulation capability: parameters that SIESTA cannot natively reproduce (such as hybrid functionals (e.g., HSE, PBE0), plane-wave cutoffs, or specific dispersion corrections) are intercepted and explicitly recorded as unmappable methodological divergences.

\subsection{Phase 4: Recomputation}
\label{sec:recomp-methods}

Candidate materials are prioritized by reporting completeness. Because relaxed coordinates are rarely deposited, initial structures are constructed from topological prototype geometries parameterized with published lattice constants, and executed with standard SIESTA configurations.

\paragraph*{Automated convergence and diagnosis.} Convergence testing is an autonomous loop whose length is decided at runtime. Following each variable-cell relaxation (target force tolerance \SI{0.02}{\electronvolt\per\angstrom}), the Convergence agent reads the output and selects from four actions (increasing the mesh cutoff, expanding the basis set, densifying the in-plane $k$-grid, or terminating execution), advancing monotonically along predefined parameter progressions (e.g., mesh cutoffs step along \SIrange{200}{900}{\rydberg} and bases along SZ, SZP, DZ, DZP, TZP). Termination requires successive completed steps $i-1$ and $i$ of a cell of $N$ atoms with total energy $E_i$ and in-plane lattice constant $a_i$ to satisfy:
\begin{equation}
\left|\frac{E_i - E_{i-1}}{N}\right| \le \varepsilon_E,
\quad
\frac{|a_i - a_{i-1}|}{a_{i-1}} \le \varepsilon_a,
\label{eq:conv}
\end{equation}
where $\varepsilon_E = \SI{5}{\milli\electronvolt}$ and $\varepsilon_a = \SI{0.5}{\percent}$. Convergence is never the agent's to declare: it is evaluated arithmetically in code before the agent is consulted, and any proposal that would make the calculation computationally cheaper is programmatically refused. The relaxation loop is constrained to a maximum of six iterations and a four-hour wall-clock budget per target. Non-converged calculations trigger an automated Diagnostician agent that parses error logs and extracts supporting failure lines; in the present campaign, all simulated targets achieved SCF convergence nominally without activating diagnostic recovery.

\paragraph*{The Critic agent and epistemic verdicts.} Before a final verdict is issued, the Critic agent selects verification checks to run from a vocabulary of eight options (symmetry, expected symmetry, calibration, convergence, units, structure file, literature claims, and literature methods). The agent chooses \emph{which} checks are warranted; the underlying code strictly controls \emph{what} data they run on. Compiling the valid check names directly into the GBNF decoding grammar structurally prevents the model from hallucinating non-existent tests.

The pipeline issues no categorical verdicts. Because starting from prototype geometries and using independent basis sets conflates methodological differences with reporting divergence, the system reports only raw physical deviations. When calculations fail convergence or leave their prototype symmetry, the Critic marks them as \emph{inconclusive} and withholds the comparison.

\paragraph*{Axis-matched comparison.} An in-plane cell is unordered, and publications do not agree on which axis is $a$: comparing label to label rather than magnitude to magnitude manufactured deviations of several per cent where the axis-matched agreement was sub-percent. Published and recomputed cells are therefore both sorted before comparison, and a comparison is formed only where the paper reports \emph{both} in-plane constants. A single reported constant cannot be assigned to an axis; pairing it with the nearer recomputed axis would select the smaller of two residuals by construction, so such targets are reported as raw deviations in the text and never plotted.


\section{Results}
\label{sec:results}

\subsection{Retrieval yield and corpus composition}
\label{sec:results-corpus}

Execution of the retrieval pipeline (16 August 2026) returned an initial candidate set, yielding a baseline corpus of \Ncorpus{} unique papers. Of the \Noa{} manuscripts with an open-access location, automated retrieval returned 403 Forbidden errors for \Nforbidden{} records, and validation stripped the remainder: 21 failed the minimum length threshold (Section~\ref{sec:corpus}), 7 were non-article-like landing pages, and 2 returned other HTTP errors. This yielded \Nfulltext{} full-text manuscripts, of which \Narxiv{} were sourced via arXiv.

During Phase 2, \Nextractedok{} of these full texts were successfully processed by the extraction agent. However, the agent subsequently classified \Nexcludednotpenta{} records as lacking a genuine pentagonal structural motif, representing a model-dependent classification rather than a deterministic filter. Excluding these non-pentagonal structures defines the final audited population of \Naudited{} papers.

Because a paper's arXiv copy and its journal version may report differently, the audited set was partitioned by publication status (\NpreprintPapers{} preprints with arXiv DOIs versus \NpublisherPapers{} journal articles with publisher DOIs) and compared against the strict reproducibility criterion. The analysis demonstrates significantly higher reporting completeness in preprints: \NpreprintComplete{} of \NpreprintPapers{} (\SI{\PctPreprintComplete}{\percent}) preprints state every required parameter, compared to \NpublisherComplete{} of \NpublisherPapers{} (\SI{\PctPublisherComplete}{\percent}) publisher versions (difference of \PreprintGapDiffPp{} percentage points, 95\% CI 10--63\,pp, Fisher's exact test, $p = \PreprintGapFisherP$). The mean parameter reporting rate is \SI{84}{\percent} for preprints versus \SI{62}{\percent} for journal articles.

This divergence is consistent with structural publishing constraints rather than disparities in scientific diligence. Journal formatting limits routinely displace methodological parameters into Supporting Information files, which the retrieval architecture deliberately excludes (Section~\ref{sec:corpus}). Consequently, the measured disparity likely reflects where parameters are physically located rather than whether they are entirely omitted by the authors, and the same caution applies to the extraction channel, since both proportions are measured through a pipeline whose misses are not known to be distributed evenly across the two publication routes.

\subsection{Local inference stability}
\label{sec:results-stability}

\begin{figure*}[t]
\centering
\includegraphics[width=\textwidth]{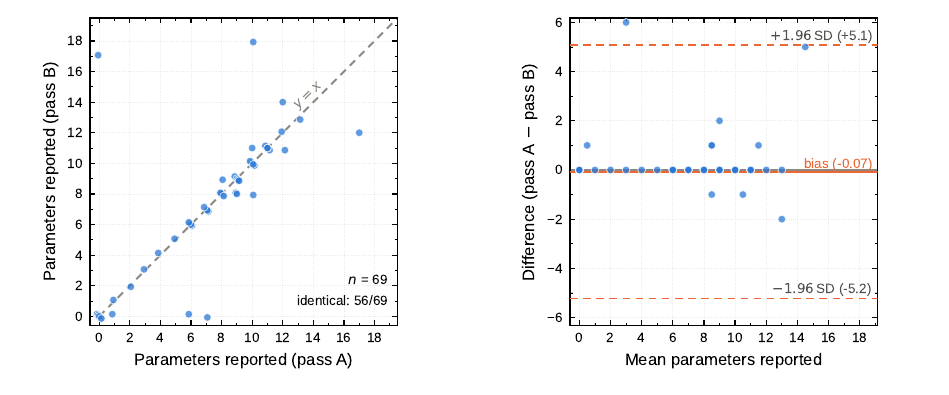}
\caption{Repeatability of extraction over two identical passes on the \SI{4}{\gibi\byte} GPU. Left: paired per-paper counts against the identity line. Right: Bland--Altman plot showing disagreement is independent of parameter count. }
\label{fig:agreement}
\end{figure*}

To quantify the stability of locally quantised inference, the pipeline was executed multiple times over the corpus. Field-level decisions remained highly stable across passes (Figure~\ref{fig:agreement}).

This instability manifests as an abrupt structural collapse rather than gradual semantic drift (Figure~\ref{fig:agreement}): a flipped boolean token forces the constrained grammar down an empty generation path. Instability results from incomplete GPU residency rather than cache precision. On a \SI{4}{\gibi\byte} GPU, the server placed between \NlayersResidentMin{} and \NlayersResidentMax{} layers on the GPU. This split varied between loads due to the uncontrolled residency condition (Section~\ref{sec:agents}). On a GPU that holds the model, three passes with an identical \texttt{q4\_0} cache are byte-identical. Consequently, a collapsed pass records zero parameters, falsely inflating the non-reporting rate.

To mitigate this hardware-induced variance, the final dataset was constructed via a grounded union of three independent passes (Table~\ref{tab:union}). This approach leverages the monotonic union defined in Section~\ref{sec:eval} to inherently recover false negatives without adjudication.

\begin{table}[tb]
\centering
\caption{Corpus-scale effect of combining passes on the \SI{4}{\gibi\byte} GPU over \Nfulltext{} documents.}
\label{tab:union}
\small
\setlength{\tabcolsep}{4pt}
\begin{tabular}{lrrr}
\toprule
 & Papers & \shortstack[r]{Fields\\reported} & \shortstack[r]{Papers\\reporting none} \\
\midrule
Pass 1         & \Nfulltext{} & \NfieldsPassOne{}   & \NzeroPassOne{}   \\
Pass 2         & \Nfulltext{} & \NfieldsPassTwo{}   & \NzeroPassTwo{}   \\
Pass 3         & \Nfulltext{} & \NfieldsPassThree{} & \NzeroPassThree{} \\
\midrule
Grounded union   & \Nfulltext{} & \textbf{\Nfields{}} & \textbf{\NzeroUnion{}} \\
\bottomrule
\end{tabular}
\end{table}

Across three passes, \Ncollapsedonce{} of \Nfulltext{} papers suffered total extraction collapse at least once. The union recovered parameters for \Nrescued{} of them, while the remaining \Ncollapsedalways{} failed in every pass (Table~\ref{tab:union}), which the pipeline records as complete non-reporting; the labelled subset does not permit these to be separated from consistent extraction failure. Overall, the union yielded \Nfields{} supported fields, recovering \Nrecovered{} fields over the best individual pass (\NfieldsPassTwo{} fields). The merge is not a plain superset: resolving plane-wave and real-space mesh cutoff collisions removes nine of the 484 fields a naive union would carry (Section~\ref{sec:eval}).

At a 24{,}576-token context window, an \texttt{f16} cache requires \SI{3.38}{\gibi\byte} of VRAM, forcing severe memory offloading on a \SI{4}{\gibi\byte} GPU. On hardware that holds both (the \SI{15}{\gibi\byte} Tesla T4), the \texttt{f16} configuration is \SI{27}{\percent} \emph{faster} than \texttt{q4\_0}. However, because \texttt{f16} cannot fit on the \SI{4}{\gibi\byte} GPU, the \texttt{q4\_0} baseline was maintained for consumer-hardware metrics.

\begin{table}[tb]
\centering
\caption{ Hardware constraints and execution profiles. Minimum VRAM includes the weight buffer, the key/value cache at the corpus's widest window (28{,}672 tokens), and the compute buffer. ``Layers'' records how many of the \NlayersTotal{} model layers fit on the GPU.}
\label{tab:configs}
\small
\setlength{\tabcolsep}{2pt}
\begin{tabular}{llrrr}
\toprule
GPU & W / KV & \shortstack[r]{VRAM\\(GiB)} & \shortstack[r]{Time\\(s)} & Layers \\
\midrule
RTX 3050 (\SI{4}{\gibi\byte}) & \texttt{Q4\_K\_M}/\texttt{q4\_0} & 3.61 & 120 & \NlayersResidentMin{}--\NlayersResidentMax{} \\
Tesla T4 (\SI{15}{\gibi\byte}) & \texttt{Q4\_K\_M}/\texttt{q4\_0} & 3.61 & 58 & 37 \\
Tesla T4 (\SI{15}{\gibi\byte}) & \texttt{Q4\_K\_M}/\texttt{f16} & 6.44 & \textbf{43} & 37 \\
Tesla T4 (\SI{15}{\gibi\byte}) & \texttt{Q8\_0}/\texttt{f16} & 8.10 & 50 & 37 \\
\bottomrule
\end{tabular}
\end{table}

Hardware holding the full model separates the two quantisation axes (Table~\ref{tab:configs}). Doubling weight precision from \texttt{Q4\_K\_M} to \texttt{Q8\_0} costs \SI{1.7}{\gibi\byte} and \SI{17}{\percent} more time. (The effect of cache quantisation on extraction accuracy is evaluated in Section~\ref{sec:results-extraction}.)

\subsection{Extraction accuracy and reporting completeness}
\label{sec:results-extraction}

\subsubsection*{Precision, recall, and grounding taxonomy}
An expert annotated \NlabelPapers{} papers against the nine required parameters, yielding \NlabelJudgements{} independent judgements blind to the pipeline's output. \NlabelStubs{} abstract-only records were excluded.
Basis size applies only to numerical-orbital codes; recall over the eight universal fields is \PctRecallEight\,\%, versus \PctRecall\,\% across all nine.

Two limits of presence-based metrics emerge. First, precision records presence, not correctness: of \NcodeChecked{} labelled packages, the extractor named \NcodeCorrect{} correctly. Remaining errors are consequential, as package identity governs the physical meaning of ambiguous numeric cutoffs (discussed below). Second, \Nconversionlost{} parameters (\PctConversionLoss\,\%) found in the PDF were destroyed in PDF-to-text conversion, which PDF-based recall would wrongly charge to the extractor.

\begin{table*}[t]
\centering
\caption{Extraction accuracy scored against expert hand labels. Accuracy is reported under both decision orientations due to extreme label skew (only \EvalPctMinority\,\% of judgements are \emph{not reported}) and because parameter \emph{absence} is the primary quantity this work measures. Matthews' correlation coefficient~\cite{matthews1975comparison} is included because a constant ``reported'' baseline scores zero.}
\label{tab:eval}
\small
\begin{tabular}{lrrrrrrr}
\toprule
 & \multicolumn{3}{c}{Reported} & \multicolumn{3}{c}{Not reported} & \\
\cmidrule(lr){2-4}\cmidrule(lr){5-7}
Configuration & P & R & F\textsubscript{1} & P & R & F\textsubscript{1} & MCC \\
\midrule
RTX 3050 \SI{4}{\gibi\byte}, \texttt{q4\_0}
  & \EvalRtxPresP & \EvalRtxPresR & \EvalRtxPresF
  & \EvalRtxAbsP & \EvalRtxAbsR & \EvalRtxAbsF & \EvalRtxMcc \\
Tesla T4 \SI{15}{\gibi\byte}, \texttt{q4\_0}
  & \EvalTfourPresP & \EvalTfourPresR & \EvalTfourPresF
  & \EvalTfourAbsP & \EvalTfourAbsR & \EvalTfourAbsF & \EvalTfourMcc \\
Tesla T4 \SI{15}{\gibi\byte}, \texttt{f16}
  & \textbf{\EvalFfsixteenPresP} & \EvalFfsixteenPresR & \EvalFfsixteenPresF
  & \EvalFfsixteenAbsP & \textbf{\EvalFfsixteenAbsR} & \textbf{\EvalFfsixteenAbsF}
  & \textbf{\EvalFfsixteenMcc} \\
\bottomrule
\end{tabular}
\end{table*}

Scored conventionally (presence as positive), every configuration exhibits high nominal performance. When evaluated against parameter absence (the primary quantity measured in this audit), precision does not exceed \EvalTfourAbsP\,\%, indicating that the majority of parameters flagged as unreported are present in the source text.

Full GPU residency, not cache precision, produces the larger effect: holding the cache at \texttt{q4\_0} on a fully-resident GPU raises recall and MCC (Table~\ref{tab:eval}). Unquantising the cache then trades presence recall for conservatism, rendering it the configuration least prone to asserting ungrounded parameters.

\begin{figure*}[t]
\centering
\includegraphics[width=\textwidth]{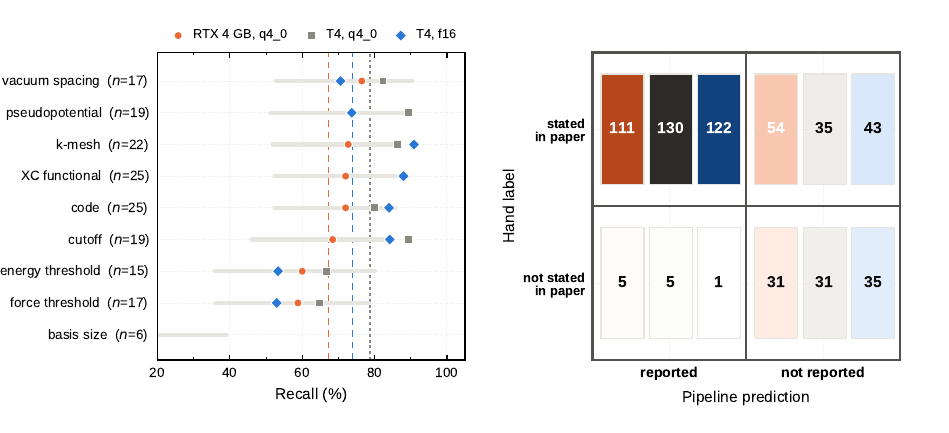}
\caption{Extractor accuracy against hand labels. Recall is shown with its 95\%
CI; \texttt{basis size} is marked because it lacks sufficient positive support to state a rate. Dashed vertical lines mark overall recall for each configuration. The confusion matrix gives the counts from which the
rates derive.}
\label{fig:accuracy}
\end{figure*}

The pipeline's baseline performance (precision, recall, and F\textsubscript{1}) is detailed in Figure~\ref{fig:accuracy}. The near-perfect precision confirms the structural efficacy of the grounding gate. In the final union dataset, \NfieldsGrounded{} of \Nfields{} asserted fields carry correctly matching textual evidence. The five false positives behind the precision score were tightly localized (\texttt{pseudopotential\_type} 2, \texttt{energy\_threshold} 2, \texttt{k\_mesh} 1). 

Mathematically, the measured non-reporting rate is defined as the sum of genuine author omissions and the extractor's false negative rate.
Writing $T$ and $O$ for the parameters genuinely reported and genuinely omitted across the
labelled judgements, and $\rho$ and $\pi$ for recall and precision, the measured
non-reporting count is
\begin{equation}
M = O + \underbrace{(1-\rho)\,T}_{\text{missed}} - \underbrace{\frac{1-\pi}{\pi}\,\rho\,T}_{\text{false positives}},
\label{eq:bound}
\end{equation}
so $M \ge O$ whenever $1-\rho \ge \rho\,(1-\pi)/\pi$, which holds here by a factor of
eleven: recall error dominates precision error by an order of magnitude. Extrapolated from the labelled subset to the full 20-field schema, this performance is of the same order as figures reported for established literature extractors such as \textsc{ComProScanner}~\cite{roy2026comproscanner}, \textsc{Cactus}~\cite{mcnaughton2024cactus}, \textsc{MiniCheck}~\cite{tang2024minicheck}, and \textsc{ChemDataExtractor}~\cite{swain2016chemdataextractor}, though on different corpora and extraction schemas. Per-field recall varies significantly: notably, \texttt{basis\_size} was never recovered (Figure~\ref{fig:accuracy}). As a strictly required field for that methodology, this represents a structural omission in extraction capability rather than an isolated variance.

Prior to the grounding gate, \PctGroundedPre\,\% of supporting sentences across the multi-pass union were accurately located in the raw text. Across the \Nvaluechecked{} extracted numeric values in the union dataset, \PctSupported\,\% were correctly derived from their quotes. The pipeline actively intercepts a specific taxonomy of five silent failure modes inherent to open-weights LLM extraction:
\begin{enumerate}
  \item \textbf{Head-truncated prompts:} The inference server silently discards the prompt head to retain the tail under constrained context (observed in 9 of 69 manuscripts).
  \item \textbf{Default-value schema hallucinations:} The model satisfies the forced schema type constraints (Section~\ref{sec:agents}) by inserting canonical placeholder literals (e.g., \num{0.0} or \num{1.0}) when parameters are unstated.
  \item \textbf{Hardware-induced structural collapse:} Partial GPU residency perturbs logits under greedy decoding, forcing the constrained grammar to silently collapse into empty JSON outputs (Section~\ref{sec:results-stability}).
  \item \textbf{Self-negating evidence:} The model sets \texttt{reported=true} while generating an evidence quote explicitly stating that the authors did not report the value (observed in \Nselfneg{} instances, \PctSelfneg\,\%).
  \item \textbf{Memorised-value substitution and plausible sweeps:} The model bypasses document retrieval by emitting canonical literature constants memorised during pretraining (e.g., substituting the ideal penta-graphene lattice constant $a=\SI{3.64}{\angstrom}$ or penta-PdSe\textsubscript{2} constant $b=\SI{5.92}{\angstrom}$ into manuscripts describing unrelated compositions), or synthesising plausible smooth numerical sweeps (e.g., stepping cohesive energies from \SI{-7.1}{\electronvolt} to \SI{-7.7}{\electronvolt}).
\end{enumerate}
The secondary value-grounding gate (Section~\ref{sec:eval}) successfully intercepted this behaviour, automatically stripping \Nclaimscleared{} ungrounded values (\Pctclaimscleared\,\% of testable valued claims) across the corpus and preventing ungrounded physical parameters from propagating into downstream simulation workflows. The two fields lacking evidence strings entirely represent a distinct, uncaught failure mode.

Testing a deterministic localiser against known evidence sentences confirms why full-context inference is required (Section~\ref{sec:eval}): keyword retrieval recovered only \SI{\PctLocaliserNarrow}{\percent} of the necessary context; doubling the retrieval window recovered \SI{\PctLocaliserWide}{\percent}, yet every parameter field continued to fail to capture critical context. The failure is fundamental to the physics text: parameters such as ``\SI{40}{Ry}'' ambiguously map to specific cutoffs depending entirely on the underlying software code. Supplying a localized context window structurally severs this dependency.

\begin{figure}[tb]
\centering
\includegraphics[width=\columnwidth]{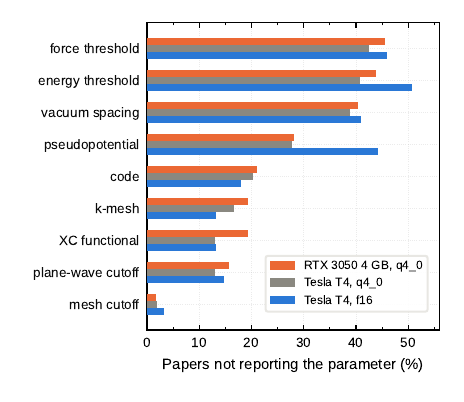}
\caption{Parameters absent from the retrieved full text, $n=\Naudited$. These rates are upper bounds on non-reporting: at \PctRecall\,\% recall a fraction of these
absences are extraction failures rather than omissions by the authors.}
\label{fig:gaps}
\end{figure}

Across the \Naudited{} papers, only \Nreproducible{} (\PctReproducible\,\%) are reproducible in principle in the main text, with a mean parameter reporting rate of \PctMeanReported\,\%. These omissions are heavily skewed toward convergence controls (Figure~\ref{fig:gaps}); notably, \SI{21.1}{\percent} of the audited papers are recorded as not naming the specific software package used for the calculations. In accordance with Equation~\ref{eq:bound}, this reflects both author omission and extractor failure: on the hand-labelled subset every paper names both its software package and its exchange--correlation functional, each of which the extractor records as missing in roughly one paper in seven. A third channel is methodological: ten of the twelve records naming no package also report no functional, pseudopotential or cutoff, a pattern absent from every paper that does name one. Such records are consistent with studies that are computational but not first-principles, and are therefore scored against parameters their methodology never defines. Paper-level completeness represents the most statistically volatile metric reported: the strict $f(p)=1$ criterion (Equation~\ref{eq:fraction}) is acutely sensitive to single-field errors, since a single false negative records a paper as incomplete and a false positive as complete, so the two error channels partly cancel. On the labelled subset the pipeline and the annotator disagreed on five of twenty-five papers, in both directions, with a net difference of one.

\begin{table*}[t]
\centering
\caption{Agent activity across the recomputation campaign. The multi-agent system was evaluated in three configurations differing only in GPU and KV-cache precision: the original RTX 3050 \SI{4}{\gibi\byte} with \texttt{q4\_0} cache (12 targets), and two fully-resident Tesla T4 \SI{15}{\gibi\byte} runs over 9 targets: one with a \texttt{q4\_0} cache, and one with an unquantised \texttt{f16} cache (pooling 17 total invocations across a repeated run). For a given target, the extraction and prompts remain identical. Only the Critic operates over multiple tool-calling rounds, and it alone collapsed in every configuration.}
\label{tab:agents}
\small
\begin{tabular*}{\textwidth}{l@{\extracolsep{\fill}}rrrrrrrrr}
\toprule
 & \multicolumn{3}{c}{RTX 3050 (\texttt{q4\_0})}
 & \multicolumn{3}{c}{Tesla T4 (\texttt{q4\_0})}
 & \multicolumn{3}{c}{Tesla T4 (\texttt{f16})} \\
\cmidrule(lr){2-4}\cmidrule(lr){5-7}\cmidrule(lr){8-10}
Agent & Calls & Collapsed & Tool calls
      & Calls & Collapsed & Tool calls
      & Calls & Collapsed & Tool calls \\
\midrule
Translator    & \NtranslatorCalls{} & 0 & --- & 9 & 0 & 1
              & 17 & 0 & 1 \\
Convergence   & \NconvCalls{} & 0 & --- & 15 & 0 & ---
              & 26 & \textbf{1} & --- \\
Diagnostician & 0 & --- & --- & 0 & --- & ---
              & 0 & --- & --- \\
Critic        & \NcriticCalls{} & \NcriticCollapse{} & \NcriticToolCalls{}
              & 9 & \textbf{3} & 16
              & 17 & \textbf{1} & 52 \\
\midrule
All agents & \NallAgentCalls{} & \NcriticCollapse{} & --- & 33 & 3 & --- & 60 & 2 & --- \\
\bottomrule
\end{tabular*}
\end{table*}

\subsection{Agent activity and the effect of cache quantisation}
\label{sec:results-agency}

The activity and stability of the four execution agents (Table~\ref{tab:agents}) were evaluated to track cache quantisation instability across roles.

\textbf{Translator.} The agent mapped parameters from \NverifRecords{} verification records into \NtranslatorSets{} distinct sets. The code-gate intercepted all \NtranslatorCalls{} invocations because the agent emitted standard literature notation (e.g., \texttt{GGA/PBE}) rather than strict parser syntax. Across a single input over ten passes, the Translator produced identical configurations nine times; the sole discrepancy (a mesh cutoff) was corrected deterministically by the convergence routine.

\textbf{Convergence.} Across \NconvCalls{} invocations, the agent uniformly selected \texttt{increase\_mesh}; because it never exercised alternative discretion, it did not alter any physical outcomes. Because loop termination is deterministically orchestrated (Section~\ref{sec:agents}), agent collapse cannot falsely declare convergence; it can only halt execution. Convergence typically required two to four steps, never reaching the imposed iteration or wall-clock limits.

\textbf{Diagnostician.} The agent was not invoked; no calculation suffered self-consistent field (SCF) or structural failure.

\textbf{Critic.} It raised 18 concerns across seven physical categories (six blocking errors, 12 warnings) over nine verification records, downgrading \NcriticChanged{} verdicts to inconclusive and identifying geometric symmetry breaks. In one instance, failing to select checks triggered a deterministic override that substituted the complete set. The Critic occasionally failed to return a schema-valid object, leaving those verdicts unreviewed. Because the verification rule set enforces an asymmetric hierarchy (Section~\ref{sec:agents}), it cannot override its own contrary evidence, and a collapse safely defaults the pipeline to the unadjusted numerical conclusion.

\textbf{Instability scales with context injection, not output length.} Collapse correlates with injected tool context rather than generation volume. The Translator safely emits a median of $\sim$2400 characters; the Critic emits $\sim$1400. However, the Critic is uniquely injected with the full JSON output of executed checks before issuing its verdict. Repeating the campaign on fully-resident hardware isolates the cause (Table~\ref{tab:agents}). The Critic collapsed in \NcriticCollapse{} of \NcriticCalls{} and 3 of 9 invocations on a quantised cache (evaluated across two GPUs at indistinguishable rates, $p = 0.66$), compared to 1 of 17 on an unquantised cache. Unlike the extraction phase, the Critic's collapse thus survives full residency. It falls roughly fivefold under unquantised cache precision, though this lacks significance either on identical targets (3 of 9 vs 1 of 9, $p = 0.58$) or pooled (6 of 22 vs 1 of 17, $p = 0.11$). The effect remains non-uniform: Convergence collapsed once in 26 \texttt{f16} invocations and never in 39 \texttt{q4\_0} invocations, while the Translator never collapsed. Consistent with context-injection fragility, the Critic issued 1.8 tool calls per invocation on the quantised run versus 3.1 unquantised, failing early in its tool loop rather than after extended generation. Cache precision, operating on dynamically injected tool context, remains the primary explanation for Critic instability, despite lacking statistical significance when pooled across roles. Its effect is confined to the agents: across the \NcacheShared{} targets run under both cache precisions, the recomputed cells differ by at most \SI{\CacheCellMax}{\angstrom}, so the electronic structure is untouched.

\subsection{Recomputation of published lattice constants}
\label{sec:results-recomp}

Complete reporting is a necessary condition for reproduction, not a sufficient one: a methods section may state every parameter and still not determine the calculation, because the relaxed coordinates are almost never deposited. Recomputation directly assesses what textual auditing cannot: whether complete methodological reporting enables successful physical simulation execution. The execution pipeline ran the top-ranked materials possessing the most complete sets of reported computational fields. Standard SIESTA input configurations, alongside the published lattice parameters and atomic coordinates, were used to start the calculations.

\begin{figure}[tb]
\centering
\includegraphics[width=\columnwidth]{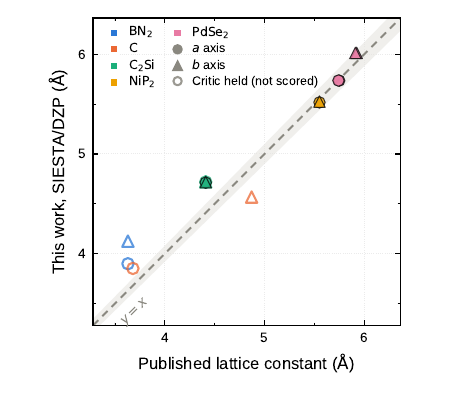}
\caption{Published versus recomputed in-plane lattice constants ($y=x$ dashed line; shaded band $\pm\SI{2}{\percent}$) across \NclaimsTierTwo{} comparisons. Circles and triangles denote the shorter lattice parameter $a$ and longer lattice parameter $b$, respectively; open markers indicate unverified targets withheld by the Critic agent due to symmetry departure~\cite{li2016penta,xu2017anisotropic,zhou2025unveiling}.}
\label{fig:parity}
\end{figure}

Figure~\ref{fig:parity} reports \NclaimsTierTwo{} matched comparisons from \NpapersTierTwo{} papers~\cite{lopez2015sigma,shao2018electronic,nandan2021performance,bravo2022topological} across \NprototypesTierTwo{} compositions, and additionally shows, faded, the targets whose relaxed cell left its prototype. The mean absolute error (MAE) is \SI{\MAEDeviation}{\angstrom} (mean absolute relative error (MARE) \MAREDeviation\,\%), median absolute deviation is \MedianDeviation\,\%, and \NwithinTwoPct{} of the \NclaimsTierTwo{} comparisons fall within \SI{2}{\percent}. The signed bias is \SI{\BiasDeviation}{\angstrom}: the recomputation is systematically larger than the published cell rather than centred on it. All four reference calculations are plane-wave (VASP, Quantum ESPRESSO), against which a DZP numerical-orbital basis is expected to overestimate, so the offset is a basis-set difference rather than a disagreement about the material. Physical agreement nonetheless varies by composition: penta-PdSe\textsubscript{2} matches closely, spanning \SIrange{0.08}{1.73}{\percent} across four comparisons, penta-NiP\textsubscript{2} yields sub-percent deviation (\SIrange{0.46}{0.61}{\percent} across two comparisons), whereas penta-C\textsubscript{2}Si deviates by $\sim$\SI{6.9}{\percent}. Two further targets are withheld because their papers report only one in-plane constant~\cite{li2022theoretical} (Section~\ref{sec:recomp-methods}). One of them exposed a compositional misattribution (substituting elemental Si for a p-SiS\textsubscript{2} claim), satisfying the textual evidence gate while specifying an incorrect material.

The Critic correctly withheld verdicts when recomputed structures left their prototype, excluding them from every deviation reported above; Figure~\ref{fig:parity} retains them as open markers. For penta-BN\textsubscript{2}, the cell broke symmetry to $b/a = 1.058$ against a tetragonal prototype; for a carbon paper, it reached $b/a = 1.1857$ (a \SI{18.6}{\percent} departure). Both triggered symmetry concerns, placing the structures on hold rather than plotting invalid comparisons.

\begin{figure*}[t]
\centering
\scalebox{1}[0.9]{\includegraphics[width=\textwidth]{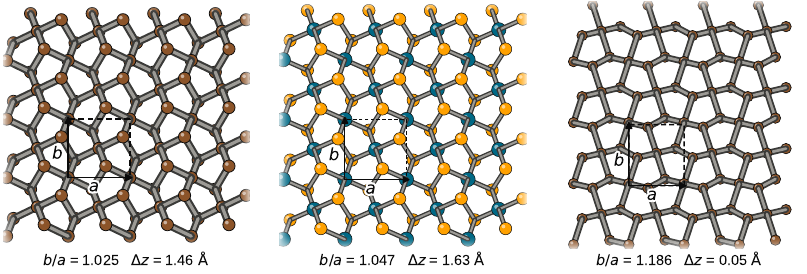}}
\caption{Visualizing conformational entrapment during variable-cell relaxation. (Left) The ideal pentagonal starting prototype. (Middle) Relaxed penta-PdSe\textsubscript{2} retains its pentagonal symmetry and proper out-of-plane buckling ($\Delta z = \SI{1.63}{\angstrom}$). (Right) The carbon target exemplifies a failure mode: the structure is trapped in a different local minimum where the $b/a$ ratio expands to 1.186, flattening the lattice ($\Delta z = \SI{0.05}{\angstrom}$) and stretching the bonds.}
\label{fig:structures}
\end{figure*}

The dominant residual error is structural rather than electronic. Because relaxed
CIF (Crystallographic Information File) coordinates are rarely deposited, structures are rebuilt from topological
prototypes, and a relaxation that settles into a different local minimum measures the
starting geometry rather than the published physics; the two symmetry holds above
exemplify this exact failure mode (Figure~\ref{fig:structures}). A perfectly reported methods section can therefore remain
irreproducible when the geometry is withheld.

Calibration against public reference geometries yielded only four usable points across two distinct pentagonal structures, falling short of the required threshold of six. No categorical verdict is therefore issued (Section~\ref{sec:recomp-methods}). Because recomputations employ independent basis sets and pseudopotentials, the raw physical deviations shown in Figure~\ref{fig:parity}, rather than categorical pass/fail verdicts, constitute the primary reported measurement.


\section{Discussion}
\label{sec:discussion}

The capability audit demonstrates a \SI{4}{\gibi\byte} \texttt{\Model{}} possesses usable semantic competence but severe structural fragility, necessitating rigid programmatic gates for transcription and data provenance. Extraction recall dominates systematic error: since the pipeline measures reporting \emph{absence}, extraction failures (false negatives) artificially inflate apparent omission rates. The \PctRecall{}\% recall ensures reported per-parameter omission rates are conservative upper bounds, measured on nine parameters at \PctPrecision\,\% precision (95\% CI \PctPrecisionLo--\PctPrecisionHi\,\%; five false positives per 116 asserted presences). Incomplete GPU residency compounds this, precluding single-pass reproducibility on \SI{4}{\gibi\byte} hardware and mandating the three-pass grounded union as a methodological necessity.

\paragraph{Defining omission within the corpus}
The pipeline strictly targets direct arXiv or open-access text, excluding separately indexed DOIs and supplementary files. Consequently, all reported rates reflect full-text presence alone. This contextualises the preprint--publisher gap (Section~\ref{sec:results-corpus}): journal conventions displace parameters into inaccessible supplementary files, meaning the gap likely measures parameter localisation rather than absolute presence. Resolving this requires dedicated per-publisher supplementary retrieval.

\paragraph{Methodological limitations}
Evidence grounding bounds fabrication but permits misattribution. For example, a p-MS\textsubscript{2} (M = Si, Ge, Pb) paper's lattice claim was stored solely under the metal, so a p-SiS\textsubscript{2} constant was attributed to elemental Si. That target was excluded from Figure~\ref{fig:parity} because the source reported only a single in-plane constant; consequently, the misattribution was identified solely through manual inspection rather than automated filtering. The pipeline cannot detect such composition mismatches; they remain identifiable solely through expert manual annotation. These labels cover nine fields across \NlabelPapers{} papers (\NlabelJudgements{} judgements) by a single expert annotator, yielding a wide \PctRecallLo--\PctRecallHi\,\% recall interval. Additionally, the \texttt{basis\_size} parameter's extreme rarity precluded statistically meaningful evaluation of its extraction recall. A related limitation attaches to corpus definition rather than to extraction: the screening flag distinguishes computational from non-computational work, not density functional theory from other computational methods. Studies employing tight-binding, classical dynamics or continuum models therefore remain in the audited population, where their absent fields are recorded as omissions rather than as evidence of a different methodology.

\paragraph{Implications for materials science reporting}
Methodological completeness does not guarantee structural reproducibility: where relaxed geometries are withheld, the rebuilt prototype rather than the published structure determines the result (Section~\ref{sec:results-recomp}). With mature deposition infrastructure (AiiDA~\cite{huber2020aiida}, NOMAD~\cite{ghiringhelli2023shared}), this barrier remains strictly conventional.

Threshold-based calibration encounters similar data scarcity: the calibration set yielded only two of the six distinct structures required before a threshold is defined, so no comparison in this campaign rests on a calibrated tolerance and no categorical verdict is issued. Moreover, independent basis sets and pseudopotentials conflate methodological divergence with basis-set incompleteness and starting-geometry entrapment. Decoupling these requires the authors' original structures. Consequently, the pipeline issues no \emph{refuted} verdicts, reporting only the raw physical deviation between published claims and independent recomputations.

\paragraph{Provenance of the reported campaign}
Two artefacts remain intentionally unreconciled: \NcriticCollapse{} unreviewed verdicts from Critic collapses, and \NverifRecords{} verification records spanning three distinct configuration fingerprints. These surface solely because outputs are stamped with their generative configuration. The capacity to detect its own systemic drift is a precondition for evaluating external literature reproducibility. The reported campaign is itself one of \NdevCampaigns{}: each defect the gates exposed invalidated every extraction that preceded its correction, and the retained logs record at least \NdevExtractions{} extractions, approximately \NdevTokens{}~million input tokens, before the corpus reported here was produced. It is that iteration, rather than the final run, which unmetered local execution makes practicable.

\section{Conclusions}

This work presents an autonomous, local framework for reproducibility auditing in computational materials science. It demonstrates an open-weights \texttt{\Model{}} model on consumer hardware can drive an end-to-end first-principles workflow. Despite severe structural fragility, such models exhibit sufficient semantic competence for complex extraction when strictly constrained by deterministic code gates. Such a model can be trusted in what it asserts, but not in what it fails to find; and that reliability depends on the conditions of its deployment, complete residency on the accelerator above all. Trust is conferred not by the model but by the constraints imposed upon it: every assertion must quote its source verbatim, and every magnitude must be located within the document before it may inform a calculation. Neither constraint is negotiable by the agent, and neither depends upon a network, a subscription, or a service that may be withdrawn.

The automated audit reveals incomplete textual reporting is not the binding reproducibility constraint. Successful parameter extraction does not guarantee structural reproducibility. Physical recomputations indicate the primary barrier to data-driven discovery is absent structural data, not omitted parameters. With relaxed coordinates rarely deposited, workflows must rely on topological prototypes. This forces conformational entrapment, conflating true methodological divergence with starting-geometry sensitivity.

Definitively adjudicating published claims requires addressing these systemic data gaps. Future pipelines must parse publisher-specific supplementary materials housing siloed structural files and build community-driven reference geometry datasets. Finally, while this implementation orchestrates density functional theory relaxations, the underlying framework remains property-agnostic. Expanding verification workflows to derived physical properties (such as elastic constants, formation energies, and electronic band gaps) will establish local open-weights agentic systems as viable instruments for continuous verification of the scientific record.



\section*{Data and code availability}
The complete pipeline source code and configuration files are available at \url{https://github.com/shambhubhandari/Agentic-Workflow} (MIT license). Raw full-text manuscripts are omitted for copyright compliance; the retrieval module provides the OpenAlex queries and repository-fallback routines to regenerate the corpus from public endpoints. The repository provides the agent-extracted records, the hand-labelled ground-truth dataset, and all SIESTA input/output files. Every output file is cryptographically stamped with its generative configuration to guarantee auditability.

\section*{Conflicts of interest}
There are no conflicts to declare.

\section*{CRediT authorship contribution statement}
\textbf{Shambhu Bhandari Sharma (S.B.S):} Conceptualization, Methodology,
Software, Formal analysis, Investigation, Data Curation, Validation,
Visualization, Writing -- original draft, Writing -- review \& editing.

\section*{Acknowledgements}
The author acknowledges Alibaba Cloud (\texttt{Qwen}), \texttt{llama.cpp}, and Ollama for enabling local, grammar-constrained execution. Bibliographic retrieval used OpenAlex (OurResearch); recomputations used SIESTA. The author thanks the developers of PyMuPDF, BeautifulSoup, and the Python ecosystem.
\bibliographystyle{rsc}
\bibliography{reference}

\end{document}